\documentclass[11pt,a4paper]{article}

\usepackage{multibib}
\usepackage[T1]{fontenc}         
\usepackage[latin1]{inputenc}
\usepackage{graphicx}
\usepackage{amsmath,amssymb}
\usepackage{bm}
\usepackage{tikz}
\usepackage{color}
\usepackage{helvet}
\usepackage{courier}
\usepackage{makeidx}
\usepackage{footmisc}
\usepackage{type1cm}         
\usepackage{textcomp}
\usepackage{subfig}
\usepackage{enumitem,xcolor}    
\usepackage{yhmath}
\usepackage{esdiff}
\usepackage{mathtools}

\DeclareCaptionLabelSeparator{colon}{. }
\usepackage[labelfont={bf},labelsep={colon}]{caption}

\makeatletter
\DeclareRobustCommand\dalemb{\mathpalette\inner@dalemb{}}
\def\inner@dalemb#1{%
  \add@dalemb#1{03}%
  \add@dalemb#1{06}%
  \square
}
\def\add@dalemb#1#2{%
  \sbox0{\scalebox{1.#2}{$#1\square$}}%
  \rlap{\lower0.#2\ht0\box0}%
}
\makeatother

\makeindex

\DeclareMathAlphabet\mathbfcal{OMS}{cmsy}{b}{n}

\begin{document}
\thispagestyle{empty}
\bibliographystyle{elsarticle-num} 

\begin{center}
\textcolor{blue}{ \Large  \bf  An alternative to the concept of continuous medium } \\
\vspace{3.mm}
{\bf Jean-Paul Caltagirone } \\
\textcolor{blue}{\texttt{ \texttt{  \small calta@ipb.fr }  } } \\
\vspace{3.mm}
{ \small Bordeaux INP, University of Bordeaux, CNRS UMR-5295, \\
        Arts et M{\'e}tiers Institute of Technology, INRAE, I2M Bordeaux, \\
        33400 Talence -- France  } 
\end{center}

\small

\textcolor{blue}{\bf Abstract}

Discrete mechanics proposes an alternative formulation of the equations of mechanics where the Navier-Stokes and Navier-Lam{\'e} equations become approximations of the equation of discrete motion. It unifies the fields of fluid and solid mechanics by extending the fields of application of these equations to all space and time scales. This article presents the essential differences induced by the abandonment of the notion of continuous medium and global frame of reference. The results of the mechanics of continuous medium validated by fluid and solid observations are not questioned. The concept of continuous medium is not invalidated, the discrete formulation proposed simply widens the spectrum of the applications of the classical equations.

The discrete equation of motion introduces several important modifications, in particular the fundamental law of the dynamics on an element of volume becomes a law of conservation of the accelerations on an edge. The acceleration considered as an absolute quantity is written as a sum of two components, one soledoidal the other irrotational according to a local orthogonal Helmholtz-Hodge decomposition. The mass is abandoned and replaced by the compression and rotation energies represented by the scalar and vectorial potentials of the acceleration. The equation of motion and all the physical parameters are expressed only with two fundamental units, those of length and time.
The essential differences between the two approaches are listed and some of them are discussed in depth. This is particularly the case with the known paradoxes of the Navier-Stokes equation or the importance of inertia for the Navier-Lam{\'e} equation.

\textcolor{blue}{\bf Keywords}

Discrete Mechanics; Conservation of Acceleration; Helmholtz-Hodge Decomposition; Navier-Stokes equations; Navier-Lam{\'e} equations, Galilean reference frame 

\vspace{-3.mm}
\begin{verbatim}
___________________________________________________________________________
\end{verbatim}
\vspace{-2.mm}
NOTE: The final publication is available at link.springer.com
\vspace{3.mm}

J.-P. Caltagirone, (2021), An alternative to the concept of continuous medium, Acta Mechanica, doi: 10.1007/s00707-021-03070-w

\vspace{-6.mm}
\begin{verbatim}
___________________________________________________________________________
\end{verbatim}

\normalsize

\section{Introduction}

The concept of continuous medium developed and used for centuries has led to theoretical predictions which are in agreement with physical observations in a large number of fields of physics in particular in mechanics \cite{Lan59} and in classical field theory including relativity \cite{Lan71b}. The analysis conducted in this article neither aims to discuss its well-founded or its sustainability and even less the results achieved.

Yet important discoveries of the last century do not seem to have impacted the formalism of solid or fluid mechanics established for three centuries and unanimously accepted by the mechanical community. This is the case for the equivalence of mass and energy established by Albert Einstein; in special relativity the variation of the mass with the velocity through the Lorentz factor made it possible to account for the limitation of the velocity of a particle to that of light; when the velocity tends towards the celerity of light the mass of the particle increases towards infinity.
 If the principle of equivalence between energy and mass is conventionally used by particle physicists, its extension to mechanics has not been achieved. In fluid mechanics in particular mass plays a central role and its conservation is one of its fundamental principles. The recently established formalism \cite{Cal19a}, \cite{Cal20a} has made it possible to abandon the notion of mass or density replaced by that of energy. The fundamental law of dynamics becomes an equality between accelerations i.e. the acceleration of the material medium or of a particle is equal to the sum of the accelerations applied to it. It is also, in a certain way, to integrate definitively the principle of equivalence of Galileo in the equations of mechanics.

Similarly, the works of Helmholtz and Hodge on the decomposition that bears their names have remained confined to a few applications in mathematics or image processing. Its most common use is probably the projection of a vector field on a zero divergence field in order to make it divergence-free \cite{Gue06}. If this decomposition applies to a vector of physics there is no reason that the acceleration escapes this rule.

Even if the notion of continuous medium has indisputable advantages, the derivation at one point, integration, mathematical analysis, ... the reduction at one point of the different quantities, variables and physical parameters, is not without posing coherence problems solved at the cost of hypotheses and approximations which limit its generalization to certain areas of physics. It is for example the introduction of fictitious forces into the equation of motion to compensate for well real forces and translate the mechanical equilibrium. It is also the fact that inertia cannot be formulated according to a Helmholtz-Hodge decomposition.
Other limitations are of a different nature, that introduced by Newton's linear law for viscosity leads to the propagation of longitudinal or transverse waves at infinite celerities; these linear laws are also present in other fields of physics, Fourier law, Fick law, ..., these are approximations of reality which deny the hyperbolic character on small scales. This has no impact in current applications but the corresponding equations are not relativistic.

This article discusses the concepts adopted over time for the equations of mechanics by noting the consequences that they imply in particular the use of global reference frame which leads to express equality on each of the components of an inertial frame of reference.
Another drawback of the continuous approach is linked to the transition to discrete which requires a spatial discretization step on the basis of numerical methodologies disconnected from physical modeling.

\section{Originality of the work} 

The first works on discrete mechanics date from about ten years ago, but they have been published only recently: \cite{Cal20a} \cite{Cal20b} \cite{Cal20c} \cite{Cal21b} \cite{Cal21d}. They address different aspects of this concept based on the existence of a local reference frame linked to a segment where intrinsic and imposed accelerations are expressed. The objective of this section is to briefly review the results presented in these first publications in order to clarify the novelty of the results presented here, which essentially concern the interest of the discrete concept compared to a continuous medium formulation.

The first contribution on the subject \cite{Cal20a} addresses the equivalence of the concepts used in fluid mechanics and those of solid mechanics, as well as their coupling, for example in Fluid-Structure Interaction. Although continuum mechanics is supposed to achieve the unification of the two domains, the Navier-Stokes and Navier-Lam{\'e} equations nonetheless present differences; the most visible one is the presence of an equation adjoining the Navier-Stokes equation, the conservation of mass, which is absent from the displacement formulation of solid mechanics. The discrete equation of motion formally unifies these two domains, with only longitudinal and transverse celerities serving to represent fluid and solid behaviors for various constitutive laws.

The discrete mechanics equation of motion is applied directly to heat transfer at small time and space scales \cite{Cal20b}. The hyperbolic nature of this equation at very small time constants naturally transforms into a diffusion equation as time increases. The paradox of Fourier's law, which results in an infinite heat flux when a temperature is applied and maintained on a surface at characteristic times of the order of $10^{-11} \: s$, is lifted by the hyperbolic behavior of the discrete equation at these time scales. In fact, the scalar and vector potentials of the equation of motion are energies per unit mass which take into account the heat fluxes in the medium.

Inertia takes a particular form in the discrete equation of motion \cite{Cal20c}. The two inertia terms $\nabla \left( | \bm v |^2 / 2 \right)$ and $\nabla \times \left( | \bm v |^2 / 2 \: \mathbf n \right)$ are, like all the other terms of the equation, the curl-free and divergence-free components of a Helmholtz-Hodge decomposition. In continuum mechanics it is not possible to transform the inertia term, $\bm v \cdot \nabla \bm v$ or $\nabla \left( | \bm v |^2 / 2 \right) - \bm v  \times \nabla \times \bm v$, into two orthogonal terms. The consequences are of the second order and appear only when the gradient or curl operators are applied directly to the equation of motion.

The article \cite{Cal21b} refers mainly to numerical simulations performed with the discrete model for two-phase flows. The existence of a local reference frame where accelerations and velocities are expressed facilitates the treatment of jump conditions at interfaces. For example, the consideration of capillary terms is also conceived from a sum of two orthogonal terms which, from the outset, eliminate the spurious currents observed when using more global capillary models. From the numerical point of view, the discrete model is very close to mimetic methods \cite{Sha96} and Discrete Exterior Calculus methods \cite{Des05}. The remarkable properties of these methods, the orthogonality of some terms, and the fact of mimicking the properties of the continuum, are fully recovered by the discrete formulation.

The article \cite{Cal21d} is presented as a return to the assumption of a continuous medium by integrating into the Navier-Stokes equation some of the results of discrete mechanics. In classical mechanics all the physical effects are expressed by its three components in a global orthonormal reference frame, whereas all the direct and induced accelerations are represented on a segment in discrete mechanics. Nevertheless, it is possible to transform some terms of the discrete equation of motion into equivalent terms in a derivative formulation of the Navier-Stokes equation. In particular, the two inertia terms are adapted to a representation on the three planes of a global reference frame. The new form of the Navier-Stokes equation thus possesses certain mathematical properties that can be exploited to study its convergence. However, the continuous medium approach still has some limitations that have been identified in this paper.

\vspace{2.mm}

The components of the discrete model presented here differ from the results obtained and published previously. What the theoretical results and previous simulations show is that, in standard situations, the results of classical mechanics are strictly those of discrete mechanics. Numerous simulations of analytical solutions (Couette, Poiseuille, etc.) show that the numerical solutions are exact to machine precision when they are polynomials of a degree less than or equal to two. Similarly, standard CFD test cases (lid-driven cavity, backward facing step flow, unsteady flow around a cylinder, two-phase flows, etc.) are reproduced with an accuracy of order two in space and time. For the vast majority of standard cases, the continuous and discrete formulations lead strictly to the same solutions.
The aim is to propose a discrete alternative to the classical continuous medium model and to show that some shortcomings and interpretations of classical mechanics are resolved by the discrete equation of motion. The differences are listed and commented on in section 4.1.

The objective is to show that the discrete formulation extends the domain of validity of its equation of motion for any physical parameter. Except for the application to heat transfer at very small time constants, the discrete model has been restricted to the field of mechanics. The hyperbolic character of the equation of motion, which naturally degenerates to a wave equation, would potentially serve to extend it to physical phenomena of different natures. In particular, the time span $dt$ is chosen according to the physics to be apprehended, from $dt = 10^{-20} s$ if it concerns the propagation of light waves to more important values, for example $dt = 10^{20}s$ to find a stationary solution of a fluid mechanics problem. It is in fact the grouping $dt \: c_l^2$ which potentially represents a pseudo-diffusion coefficient of the phenomenon that governs the propagation of longitudinal waves. The grouping $dt \: c_t^2$ governs the propagation of transverse waves; this quantity is highlighted in the paradox obtained (section 4.3) with the Navier-Stokes model for the startup Couette flow problem at small time constants. The discrete equation of motion allows us to find a solution where the transverse waves are first propagated in the medium before turning into a diffusion problem for larger time constants.

The compressible or incompressible character of a flow is not only imposed by the compressibility of the medium or by the variations of its density. For example, water is a medium which is not very compressible but which transmits acoustic waves at low time constants. It is still the grouping $dt \: c_l^2 \: \nabla \cdot \bm v$ that defines the importance of compressive effects. The compressible flow in a Sod tube (section 4.4) has a theoretical solution obtained from Euler's equations and energy in the case of a perfect gas. This solution is reproduced with good accuracy with the discrete model. The interest of this last formulation is that it abandons the notion of density by fixing the scalar potential at $\phi = p / \rho^{\gamma}$, where $p$ is the pressure, $\rho$ the density and $\gamma$ the ratio of the specific heats. The constitutive laws are thus dissociated from the vector equation. In this case, where the flow is isentropic, the temperature is not required even if it can be calculated {\it a posteriori}. The continuous compressible model contains several variables - velocity or momentum, pressure, temperature, density - but not all of them are independent. The sequential treatment of the Euler equations with a state law potentially involves overlaps. In discrete mechanics, only the divergence of the velocity allows one to update the energy $\phi$.

\vspace{2.mm}

The first publications show that the discrete equation of motion preserves the properties and solutions of the equations of continuous medium mechanics. The limitations of the classical equations observed for extreme physical phenomena or wider domains of validity are not so much due to the equations themselves as to the notion of continuous medium itself. This concept can be replaced by the notion of discrete medium where the conservation of acceleration on a segment is adopted as a postulate by discrete mechanics. Of course, the derivation at a point, the integration and the mathematical analysis must be adapted to the discrete medium, but the differential geometry and the exterior calculus seem to be sufficient to define the framework of discrete mechanics. This opens up new perspectives by adopting concepts from electromagnetism where direct and induced currents can be expressed from scalar and vector potentials. While the principles of mechanics are essentially associated with the notion of balance in an elementary volume to express the conservation of momentum, mass and energy, other domains of physics use different concepts based on line integrals and fluxes. The reduction of all the variable quantities and physical characteristics of the continuous medium concept to a point leads to the abandonment of the idea of direction, which then forces the reconstruction of a representation in a three-dimensional space using a global reference frame. The point of view of discrete mechanics is to keep the direction fixed by a segment of finite length; it is always possible to make the length of this segment tend towards zero in a homothetic reduction in order to derive a local equation of motion.


\section{Discrete formulation}

\begin{figure}[!ht]
\begin{center}
\includegraphics[width=5.cm]{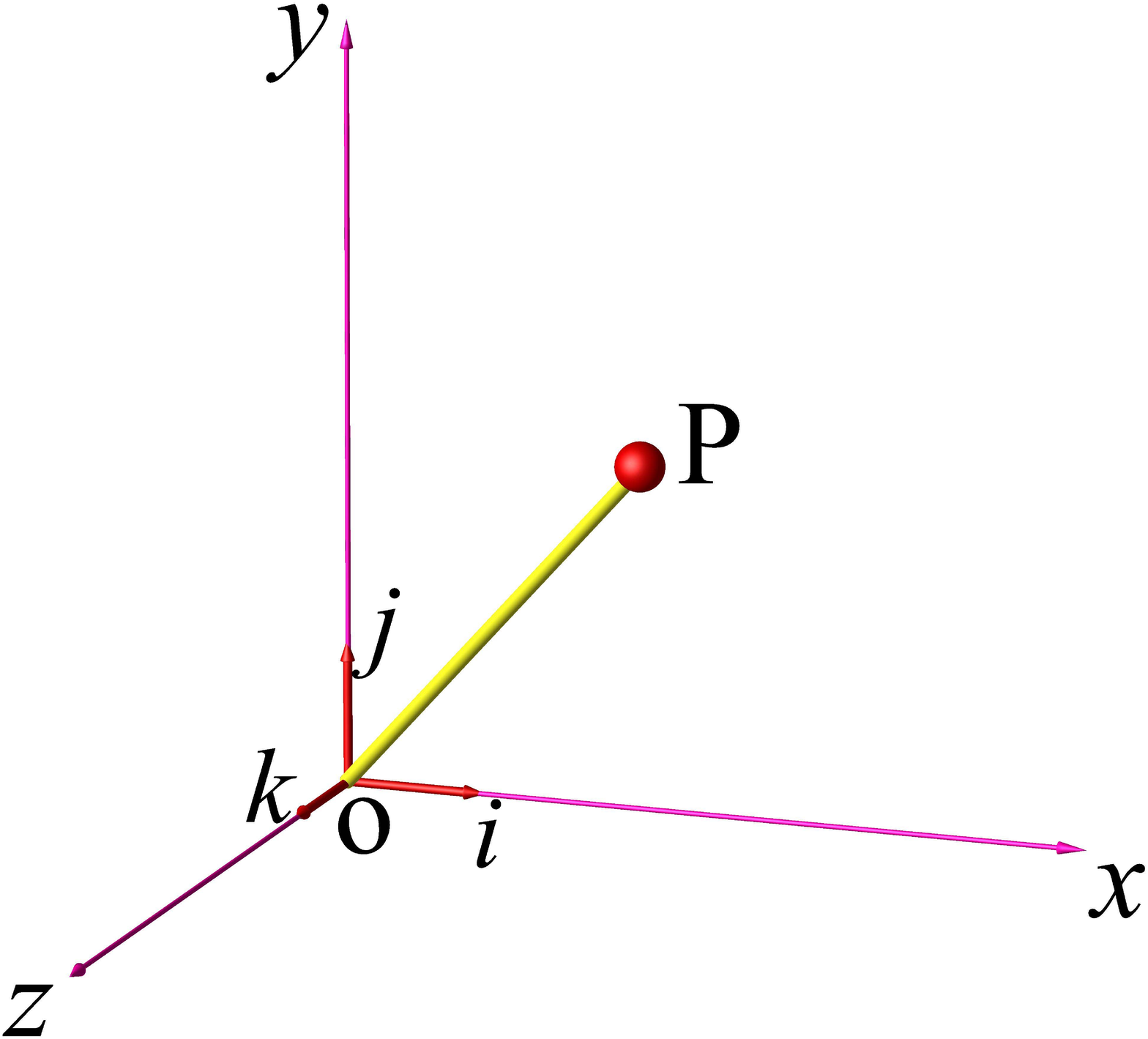}
\hspace{10.mm}
\includegraphics[width=4.cm]{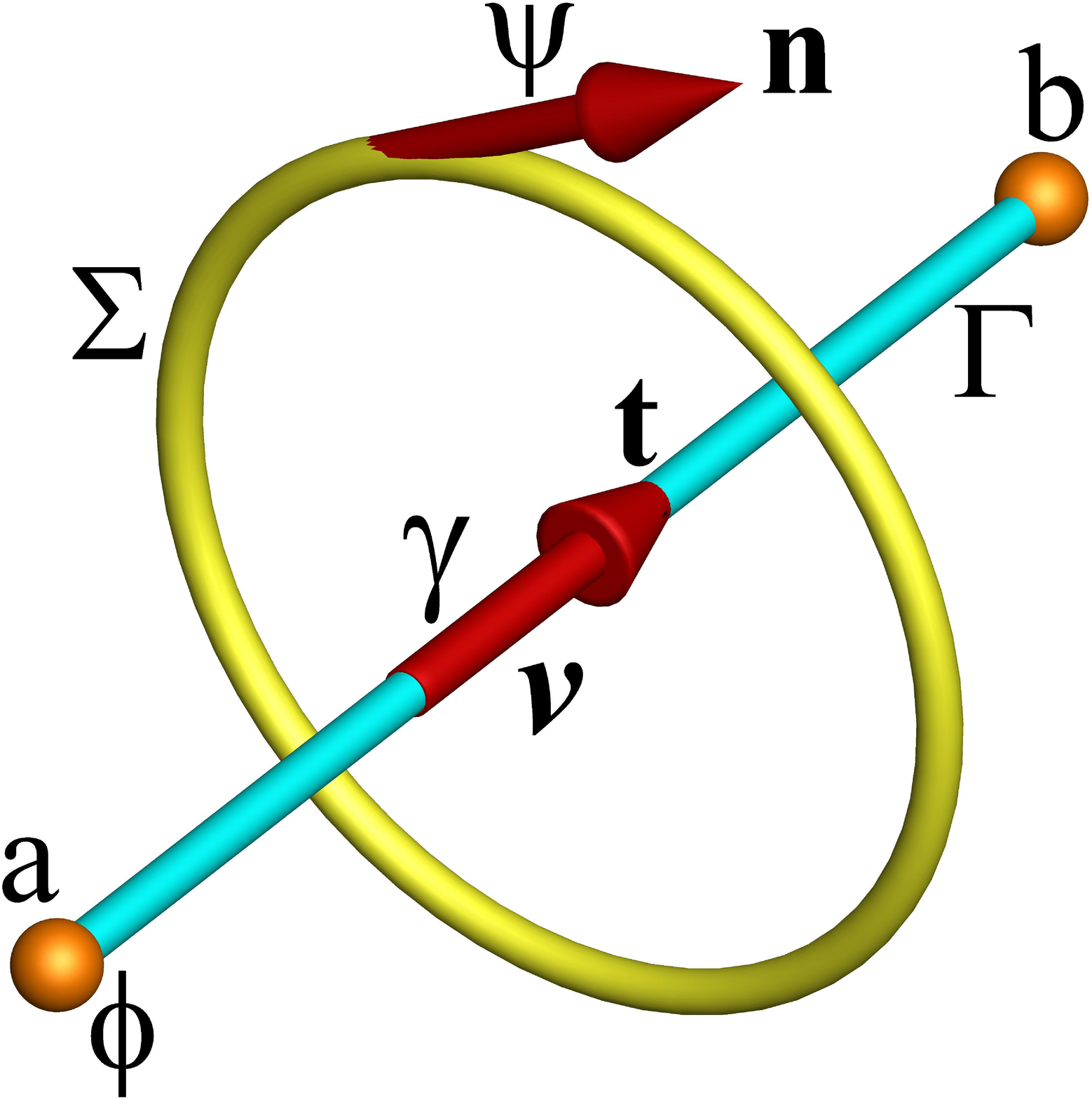} \\
\vspace{-5.mm}
(a) \hspace{70.mm}  (b)
\caption{ (a) Cartesian frame of reference and (b) discrete local geometric topology   }
\label{primdual}
\end{center}
\end{figure}

The notion of continuous medium is based on the contraction at a point of all the quantities; the figure (\ref{primdual}a) represents an inertial frame of reference $(Oxyz)$ where the point $P(x, y, z)$ carries these scalar or vector quantities. In order to reintroduce a direction and an orientation it is necessary to assign to the same vector quantity scalars attached to the axes of unit vectors $(\bm i, \bm j, \bm k)$. These components of a vector or a tensor make it possible to express this quantity at the point $P$. A null vector will thus be defined by the nullity of its three components.

The concept of discrete mechanics is based on a different vision of space and time where the reduction in a point of the variables and physical properties is not possible. The location of the material medium or of a particle in space is not known, not even in relation to an inertial frame of reference. The figure (\ref{primdual}b) represents an elementary geometric topology irreducible at a point defined by a rectilinear segment $\Gamma$ of extremity $a$ and $b$ and of length $d$. The time or rather the elapsed time $dt$ is in turn defined by the quantity $dt = d / v$ where $v$ is a constant velocity on the segment $\Gamma$. As the segment is rectilinear the velocity of the material medium or of the particle can only be lower than the celerity, $v <c_l$ where $c_l$ is the longitudinal celerity of the medium. Thus the velocity vector $\bm v$ on the edge is only one component of the velocity of the medium which will remain an unknown not necessary for the description of the motion. Remote interactions will be treated from cause to effect, information will flow from $a$ to $b$ and from $b$ to the other points of geometric topology through the wave propagation defined by $c_l$. The principle of relativity of Galileo is shown here by the fact that the velocity $\bm v$ on the segment is known only by its value $\bm v^o$ at the instant $t^o$, $\bm v = \bm v^o + dt \: \bm \gamma $ where the acceleration $\bm \gamma$ is also defined as a constant quantity on the segment. If a movement is not supersonic it cannot become it on a rectilinear trajectory. It is important not to confuse velocity and celerity; velocity is a quantity associated with a motion and celerity is a property of the medium, they should not be compared {\it a priori}.

Discrete mechanics postulates that the acceleration $\bm \gamma$ is an absolute conservative quantity on the segment $\Gamma$ i.e. the acceleration of the material medium or of the particle is equal to the sum of all the accelerations which are applied to it . This postulate is not a principle of classical mechanics nor indeed of relativistic mechanics. Of course this question will not be answered here.

The acceleration $\bm \gamma$ can only be the sum of a direct acceleration and an induced acceleration. This concept comes from electromagnetism: an electric current of intensity $\bm v$ flowing in a conductor $\Gamma$ sees its value increasing if the potential difference at the ends $- (\phi_b - \phi_a) / d$ also increases over time, this is direct acceleration. The second possibility to vary the velocity in the conductor consists in using induction i.e. circulating a current in a coil $\Sigma$ thus producing an induced acceleration in $\Gamma$. The current is induced by the equivalent of a magnetic field $\bm \psi$ produced by the circulation of an electric field on the contour $\Sigma$.
Thus the fundamental law of mechanics is written $\bm \gamma = - \nabla \phi + \nabla \times \bm \psi$.

The concepts of continuous medium mechanics attached to the inertial frame of reference of the figure (\ref{primdual}a) are therefore very different from those defined in discrete mechanics defined by the local frame of reference of the figure (\ref{primdual}b).

The principle of conservation of acceleration leads to:  
\begin{eqnarray}
\displaystyle{  \int_{\Gamma} \bm \gamma \cdot \mathbf t \: dl = - \int_{\Gamma} \nabla \phi \cdot \mathbf t \: dl + \int_{\Gamma} \nabla \times \bm \psi \cdot \mathbf t \: dl } 
\label{principia}
\end{eqnarray}
where $\phi$ is the scalar potential of $\bm \gamma$ defined on points $a$ and $b$ and $\bm \psi$ is the vector potential of the acceleration carried by the unit vector $\mathbf n$. Details of the physical derivation can be found in \cite{Cal19a} and \cite{Cal20a}; this leads to the equation of motion:
\begin{eqnarray}
\left\{
\begin{array}{llllll}
\displaystyle{ \frac{d \bm v}{d t} = - \nabla \left( \phi^o  - dt \: c_l^2 \:  \nabla \cdot \bm v   \right)  + \: \nabla \times \left( \bm \psi^o - dt \:  c_t^2 \: \nabla \times \bm v \right) + \bm h_s  } \\  \\
\displaystyle{ \alpha_l \: \phi^o - dt \: c_l^2 \:  \nabla \cdot \bm v  \longmapsto \phi^o  } \\ \\
\displaystyle{  \alpha_t \: \bm \psi^o - dt \: c_t^2 \: \nabla \times \bm v  \longmapsto \bm \psi^o }
\end{array}
\right.
\label{discrete}
\end{eqnarray}
where $\bm h_s$ represents a source term corresponding to an imposed acceleration. The quantities $c_l$ and $c_t$ are longitudinal and transversal sound celerities of medium. The parameters $\alpha_l$ and $\alpha_t$ are the attenuation factors of the longitudinal and transverse waves. These four physical parameters must be simply known even if they depend on the variables themselves.
The potentials $\phi^o$ and $\bm \psi^o$ are those describing the mechanical equilibrium at the time $t^o$. The notation $\longmapsto$ corresponds to an explicit update of the quantity at the time $t$.  

The principle of inertia applied to a rectilinear motion on a edge leads to a velocity that is independent of time when no acceleration is applied to the particle or to the material medium. 
In discrete mechanics inertia has a specific form which cannot be deduced from one of those from the continuous medium, $\bm v \cdot  \nabla \bm v$, $\nabla \cdot ( \bm v \otimes \bm v) - \bm v \: \nabla \cdot \bm v$ or $\nabla (| \bm v |^2 / 2) - \bm v \times \nabla \times \bm v$ ; the material derivative reads:
\begin{eqnarray}
\displaystyle{ \frac{d \bm v}{d t} \equiv  \frac{\partial \bm v}{\partial t} +  \nabla \left( \frac{1}{2} \:  | \bm v |^2 \right) - \nabla \times \left( \frac{1}{2} \: | \bm v |^2  \: \mathbf n \right) }
\label{inertie}
\end{eqnarray}

It is possible to integrate the first term of inertia in the second member of the equation (\ref{discrete}) to reveal the Bernoulli equilibrium potential, $\phi^o_B = \phi^o + | \bm v |^2/2 $.

In a continuous medium, inertia is written in different equivalent forms, including that which shows the Lamb vector $\mathcal L = - \bm v \times \nabla \times \bm v$ \cite{Ham08}; whatever the form used the result of the divergence of the inertial terms is the same. If we now compare the result of the divergence of the material derivative with that obtained in discrete mechanics we have:
\begin{eqnarray}
\left\{
\begin{array}{llllll}
\displaystyle{ {\rm C.M. } \:\:\:\:\: \nabla \cdot \left( \frac{d \bm v}{d t} \right)  =  \frac{\partial ( \nabla \cdot \bm v) }{\partial t}  + \nabla^2 \left( \frac{1}{2} \:  | \bm v |^2  \right) + \nabla \cdot  \mathcal L  } \\  \\
\displaystyle{  {\rm D.M. } \:\:\:\:\: \nabla \cdot \left( \frac{d \bm v}{d t} \right)  = \frac{\partial ( \nabla \cdot \bm v) }{\partial t}  + \nabla^2  \left( \frac{1}{2} \:  | \bm v |^2 \right)  }
\end{array}
\right.
\label{divaccel}
\end{eqnarray}

As the Lamb vector is not a curl its divergence equal to $\nabla \cdot \mathcal L = \bm v \cdot \nabla \times (\nabla \times \bm v) - (\nabla \times \bm v)^2$ where the first term is called flexion product and the second the negative enstrophy; $\nabla \cdot \mathcal L$ is therefore not zero even for a divergence-free velocity. There remain terms which can be expressed as a function of the second invariant $I_2$ of the tensor $\nabla \bm v$. This first important difference between the Navier-Stokes equation and the discrete formulation is directly usable for the projection methods \cite{Gue06} which consists in obtaining a velocity field with zero divergence from a starting from a prediction. of the non-zero divergence field.

\section{Continuum Mechanics vs. Discrete Mechanics }

\subsection {The main differences}

Before going into details of certain differences between the continuous and discrete formulations, it is possible to establish a non-exhaustive list of the properties of each of them.
For the continuous approach we can note:
\begin{itemize}[label = \textcolor{blue}{\textbullet}]
\vspace {1.mm}
\item the need to add mass conservation; the Navier-Stokes equation cannot be solved without associating this conservation law with it, but this is not the case with the Navier-Lam{\'e} equation which is autonomous. Although being from the same continuum mechanics the two formulations are different;
\item the value of the compression viscosity is not defined for fluids and the Stokes hypothesis is wrong, $3 \: \lambda + 2 \: \mu \ne 0$ \cite{Gad95b}, \cite{Raj13},
\item the form of the inertia terms due to the use of an inertial frame of reference and the need to express the mechanical equilibrium by component leads to fictitious forces in the equation of motion
\item the divergence of the material derivative of the velocity, in particular of the Lamb vector makes appear additional non-zero terms even for a divergence-free velocity \cite{Cal20a};
\item a longitudinal disturbance in a viscous Newtonian fluid produces a tansversal diffusion which extends instantly to infinity;
\item viscous dissipation is absent from the Navier-Stokes equation which induces a non-conservation of mechanical energy.
\end {itemize}

For the discrete approach we can emphasize that:
\begin{itemize}[label = \textcolor{blue}{\textbullet}]
\vspace{1.mm}
\item mass is useless in the equation of motion; the Weak Equivalence Principle eliminates this for accelerations linked to gravitation but also for all other accelerations applied to the material medium or to a particle; mass is replaced by energy because the potentials $\phi^o$ and $\bm \psi^o$ are respectively energies of compression and shear;
\item the physical parameters, the longitudinal and transverse velocities are perfectly defined in all media (solid, fluid, vacuum) for all propagation phenomena (gravitational wave $ c_h = \sqrt {g \: h} $, sound waves $ c_l = \gamma / \sqrt {\rho \: \chi_T} $, light $ c_0 $);
\item energy conservation is strictly ensured thanks in particular to the introduction of wave attenuation factors or a viscous dissipation term in motion equation;
\item the inertia breaks down itself into a curl-free component and another divergence-free; the divergence of the material derivative makes this last term disappear;
\item even if the question of an equation of the relativistic motion is not treated here, it appears that  the  discrete equation of motion can be written in the form of two d'Alembertians (wave operator) where $c_l$ becomes $c_0$, the celerity of light in a vacuum. The velocity of the material medium or of the particle subjected to a constant acceleration tends towards the celerity of the medium \cite{Cal19c}. This is as true for a hydraulic jump as for a straight shock wave in fluid;
\item Noether's theorem \cite {Noe11} shows that the invariances of the equation of motion by translation and by rotation lead to the conservation of the conservation of momentum and angular momentum.
Energy conservation is ensured by the invariance of the time translation (homogeneity of time).
\end{itemize}

\subsection{Stokes assumption}

Various authors have questioned the validity of Stokes' assumption in particular M. Gad-El-Hak ''Is the second coefficient of viscosity equal to the negative two-third of the dynamic coefficient of viscosity?'' \cite{Gad95b} who concludes in the negative just like K.R. Rajagopal: `` The Stokes assumption is not valid for any fluid, and this includes monatomic gases '' \cite {Raj13}. An alternative formulation of the Stokes hypothesis is given by Buresti \cite {Bur15}.
The controversies on this subject since Truesdell \cite {Tru54} are numerous and the subject is not closed. These authors bring important arguments related to the form of the stress tensor but other ways exist.
L. Landau \cite{Lan59} already proposed a characteristic time to express the second viscosity coefficient in terms of the frequency.  Several authors have measured the attenuation of sound in a medium, for example with acoustic methods, in order to determine the compression viscosity $\lambda$  \cite{Hol11} \cite{Yua13}. The measured values are always greater than the dynamic viscosity $\mu$, but the ratio is often only a few units.  But if we compare the shear-rotation viscosity of water $\mu \approx 10^{-3}$ to $dt/ \chi_T$ with $dt = 10^{-3}$, we find a ratio of the order of $10^9$! The results are the same with air. In fluids, the compression viscosity coefficient $\lambda$, even if it existed, would not influence the propagation nor the attenuation of waves.

So why are the Navier-Stokes equations representative of the phenomena observed in Newtonian fluids in such a wide variety of applications?
In practice, the viscous effects of the Navier-Stokes equations are consistent with the phenomena observed in reference experiments. The fact that $\lambda$ does not have a well-defined value, determined only by the inequality $3 \: \lambda + 2 \: \mu \geq 0$, is compensated by other equations, such as conservation of mass and state laws. Applying these additional relations makes it possible to reestablish the conservation of mass and calculate the pressure.

The equations of classical mechanics in fact only correspond to instantaneous memory models and time dependence is sometimes only ensured through the constitutive equations. Newton's linear law connecting the constraint and the orthogonal velocity gradient, $ \bm \tau = \mu \: du / dy $, has no reference to time, it is assumed to be valid at all time scales. Except it is clear that it induces, taking into account the parabolic nature of the Stokes equation, a paradox with small time constants since a perturbation at a point extends instantaneously to infinity.

The discrete motion equation does not include any constitutive equations for the viscous stress tensor. The only physical parameters are the celerities $c_l$ and $c_t$ perfectly measurable in all media (fluid, solid, vacuum); they can depend on other variables but must be known simply on each point of the geometric topology. The hyperbolic form of the equation (\ref{discrete}) leads to translate the propagation of longitudinal and transverse waves at small scales in time and to find the classical diffusion of the momentum at large scales in time. This equation is valid at all time and space scales.

It is however possible to link the Lam{\'e} coefficients $(\lambda, \mu) $ to the celerities $c_l, c_t$. The definitions $(\lambda + 2 \: \mu) / \rho = dt \: c_l^2$ and $\mu / \rho = dt \: c_t^2$ are exactly the same as for solids. Only the observation time between two mechanical equilibria $ dt $ ensures the link between the displacement and the velocity $\bm u = \bm u^o + dt \: \bm v$. This notion of observation time is essential for understanding the behavior of different media under mechanical stresses. For example, in everyday language, water which is known to be rather an incompressible medium propagates acoustic waves well with a celerity $c_l \approx 1500 \: m \: s^{-1}$, it is therefore a compressible medium for stresses with low time constants. Thus the incompressibility of a movement is not linked to the only celerity of the waves but depends closely on the time constant $ dt $ with $\nabla \cdot \bm v \approx 1 / (dt \: cl^2 )$.
For fluids the propagation of longitudinal waves is only slowly attenuated $\alpha_l \approx 1$ especially in liquids unlike transverse waves which are attenuated on very weak time constants, theoretically $\alpha_t = 0$ for a Newtonian fluid. Although it is possible to carry out simulations taking into account the transverse propagation it is preferable to replace $dt \: c_l^2$ by $\nu$ the kinematic viscosity at long times.

Thus, in discrete mechanics, there does not persist any inconsistency whatever the compressible or incompressible motions, the phenomena naturally depend on the celerities and the time of observation.

\subsection{Transverse diffusion paradox}

Let us consider the very simple physical problem of a fluid initially at rest in the half-space $ y \ge 0 $ entrained at the instant $ t = 0 $ by a solid wall positioned in $ y = 0 $ at the velocity $ u = \bm v \cdot \mathbf e_x = V_0 $ with $ V_0 = 1 $. The theoretical solution of this classical problem is obtained by the analytical resolution of the Stokes or Navier-Stokes equation without the inertial terms:\begin{eqnarray}
\left\{
\begin{array}{llllll}
\displaystyle{ \frac{\partial u}{\partial t}  - \nu \: \frac{\partial^2 u}{\partial y^2} = 0 } \\ \\
\displaystyle{ u(y,0) =  0  } \\ \\
\displaystyle{ u(y,t) =  1  } \\ \\
\displaystyle{ u(\infty,t) =  0  } 
\end{array}
\right.
\label{diffus}
\end{eqnarray}

The solution to this problem is obtained by noting that there is a self-similarity variable $ \eta $ grouping the time and space variables $ \eta = y / (2 \: \sqrt {\nu \: t}) $. Integration gives:
\begin{eqnarray}
 u(y) = erfc \left(  \eta \right) = 1 - \frac{2}{\sqrt{\pi}} \int_0^{\eta} e^{- z^2} dz  
\label{diffusol}
\end{eqnarray}

The tangential constraint $\tau_w = \nu \: du / dy = - (2 / \sqrt {\pi}) \: exp (- \eta ^ 2) \: (2 / \sqrt {\nu \: t})$ takes an infinite value when $ t \rightarrow 0 $ in $ y = 0 $. This artifact relating to Newton's law is due to the fact that, implicitly, the value of the celerity of waves is infinite. We find of course the same behavior for the Fourier law with weak time constants.

We can easily verify that the solution (\ref{diffusol}) is the same as that obtained analytically by the discrete equation (\ref{enertrans}) if we adopt the transformation $ dt \: c_t ^ 2 = \nu $ and $ \bm \psi^o = 0$. At large time constants, when the transverse waves have dissipated we find the classical diffusion solution.
\begin {figure} [! ht]
\begin {center}
\includegraphics [width = 8.cm] {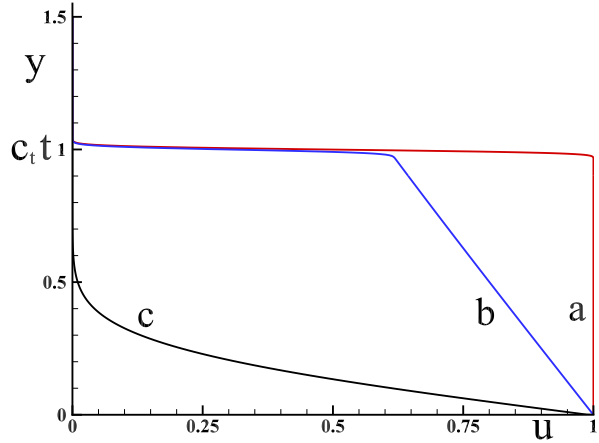}
\caption {Propagation of a transverse wave in an isotropic medium of celerity $ c_t = 1 $; the time is chosen to obtain a wavefront at a height $ y = c_t \: t = 1 $; evolution of the axial velocity $ u (y) $ for (a) $ \alpha_t = 1 $, (b) $ \alpha_t = 0.9999 $ and (c) $ \alpha_t = 0.995 $. }
\label{disnewton}
\end{center}
\end{figure}

In the general case the solution is given by the equation of motion limited to the only transverse propagation:
\begin{eqnarray}
\left\{
\begin{array}{llllll}
\displaystyle{  \frac{d \bm v}{d t} = \nabla \times \left( \bm \psi^o  - dt \: c_t^2 \:  \nabla \times \bm v  \right)  } \\ \\
\displaystyle{ \alpha_t \: \bm \psi^o - dt \: c_t^2 \:  \nabla \times \bm v  \longmapsto \bm \psi^o  } 
\end{array}
\right.
\label{enertrans}
\end{eqnarray}

This equation is solved in the upper half-space $y \ge 0$ with the condition $u(0, t) = 1$ in a dimension of space; in two dimensions of space $(x, y)$ with periodic conditions in $ x $ the solution is the same. According to the values of the attenuation coefficient of transverse waves $ \alpha_t $ the solution presents different behaviors, the spatial evolutions of the solutions are presented on the figure (\ref {disnewton}) for three values of the coefficient $\alpha_t$. It should be noted that, given the recent origin of the physical model, the values of this coefficient are not known; it is a physical parameter which must be measured experimentally or determined from the measures of attenuation in the various media. We know of course that the value $ \alpha_t = 1 $ corresponds to a propagation without dissipation of the waves and that the value $ \alpha_t = 0 $ instantaneously dissipates the transverse waves, it is the case of a purely Newtonian fluid. According to the values $\alpha_t$ we obtain:
\begin{itemize}[label = \textcolor {blue} {\textbullet}]
\item (a) - without diffusion of the transverse waves $ (\alpha_t = 1) $ the solution is hyperbolic and the velocity of diffusion of the momentum following $ y $ is limited by the celerity, $ y <c_t \: t $;
\item (b) - with a significant transverse diffusion the velocity is reduced in the fluid but the propagation front is always limited to $ y = c_t \: t $;
\item (c) - when the transverse diffusion is dominant we find the solution given by Newton's law (\ref{diffusol}).
\end {itemize}
In all cases the one-dimensional transverse wave cannot be faster than the celerity. In the classical model of Newton the wave propagates with an infinite celerity and, moreover, the shear stress is infinite at the initial time.

It is possible to model the transverse scattering at small spatial scales by a viscous diffusion term whose form $ - \sigma \: \bm v $ is identical to that of the Darcy equation $- (\nu / K) \: \bm v$ where $K$ is the intrinsic permeability of the porous medium. As $K$ is expressed in $m^2$ and $\nu$ in $m^2 \: s^{-1}$ mobility $ \sigma $ is in $s^{-1}$. Its inverse $\tau$ is the time constant of dissipation of transverse waves in a fluid whose order of magnitude is $\tau = 1 / \sigma \approx 10^{-11} - 10^{-10} \: s$; for example, for air $\tau = \nu / c_l^2 \approx 1.57 \: 10^{-5} / 347^2 = 1.30 \: 10^{-10} \: s$.

Thus there is an obvious link between the form of the term of diffusion of the momentum of the Navier-Stokes equation and the more general form of the equation of the discrete movement; this naturally degenerates to the large time constants where $dt \: c_t^2 \rightarrow \nu$. We can consider that the equation of discrete motion (\ref{discrete}) is a generalization of the Navier-Stokes equation.

\subsection{Sod shock tube problem}

The propagation of longitudinal waves is well understood by the equations of the mechanics of continuous media in their compressible formulations (Euler, compressible Navier-Stokes).
Traditionally, the one-dimensional Riemann problem of the flow in a shock tube is solved numerically from systems of equations phrased in terms of conservative variables using a suitable method (Lax-Wendroff, Osher, Van Leer, Roe, McCormack, discontinuous Galerkin, etc .). Here, we shall test the discrete non-conservative model on the case of a shock-type discontinuity.
Consider the 1D problem of a channel that is closed at both ends and separated into two compartments by a membrane. The downstream pressure is held at $ p_R $, whereas the upstream pressure is increased until the diaphragm bursts; the pressure is then equal to $ p_L $. The problem is known as Sod flow \cite {Sod78} - the one-dimensional Riemann problem of the flow of a non-viscous ideal gas. We choose the following initial conditions: $ p_L = 1 $, $ \rho_L = 1 $, $ p_R = 0.1 $, $ \rho_R = 0.125 $, $ u_L = u_R = 0 $.

Before examining the possible differences with the discrete mechanics let us write the alternative equation of the equations of Euler by simply removing the terms in dual curl of diffusion and inertia: 
\begin{eqnarray}
 \left\{  
\begin{array}{llllll}
\displaystyle{  \frac{\partial   {\bm v } }{\partial  t  }   =   - \nabla \left(  \phi^{o} -  dt \: \phi^o  \: \nabla \cdot \bm v  + \frac{| \bm v |^2}{2} \right)  } \\ \\ 
\displaystyle{  \phi^o - dt \: \phi^o \:  \nabla \cdot \bm v  + \frac{| \bm v |^2}{2}   \longmapsto \phi^o  }
\end{array}
\right.
\label{euler}
\end{eqnarray}

The equivalence between mass and energy allows us to abandon the density variable $\rho$ present in Euler's equation; This is already implicitly present in the scalar potential $\phi^o$. Furthermore the temperature is linked to the pressure $p$ and to the density through the equation of state, not necessary in the discrete formulation. The classical formulation of Euler equations $(p, \rho, T, \bm v)$ is reduced to only the quantities $(\phi^o, \bm v)$. We show \cite{Cal19a} that the compression energy $\phi^o$ is equal to the square root of celerity, $\phi^o = c_l^2 = p / \rho^{\gamma}$ for an ideal gas. This autonomous equation ensures perfect energy conservation. For compressible flows with shocks, the equation of movement is also expressed with only two fundamental units, those of length and time.

\begin{figure}[!ht]
\begin{center}
\includegraphics[width=5.5cm]{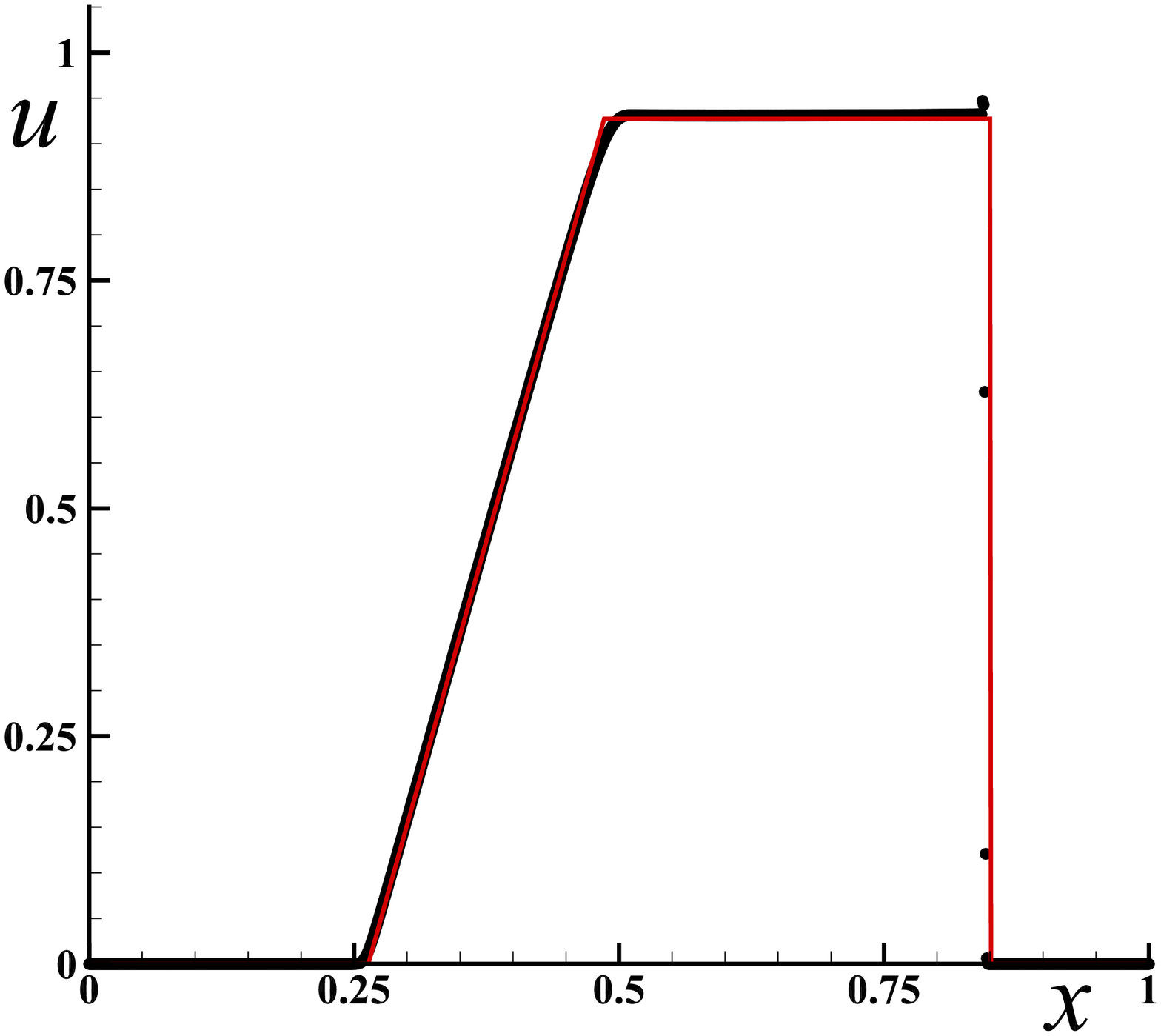}
\includegraphics[width=5.5cm]{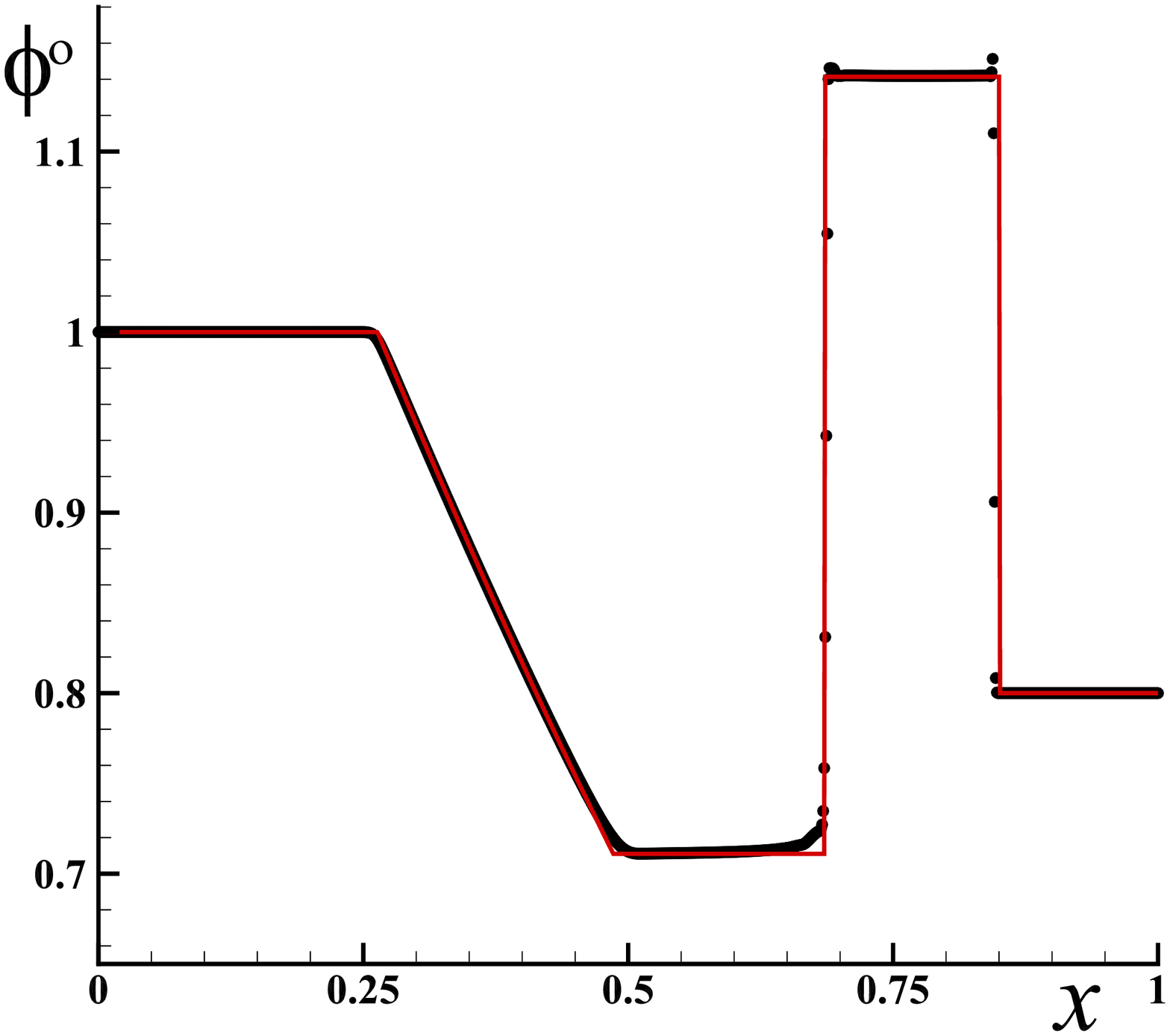}
\caption{Sod shock tube: evolution along the axis of the tube of  velocity $\bm u$ and the potential $\phi^o$;  the time step is $\delta t = 2 \cdot 10^{-5}$ and solution given at $t =  0.2$. The theoretical solution is shown as a solid line, the dotted line shows the results of the model with $N = 1000$ discretization points spanning the whole domain. Numerical results have been taken from those used in the reference \cite{Cal19a}.}
\label{tubsod}
\end{center}
\end{figure}

The results of the model are shown in figure \ref {tubsod} at time $ t = 0.2 $. The results of the simulation on the Sod shock tube problem are extracted from the reference \cite {Cal19a} and the different quantities were expressed as a function of the energy $\phi^o$ only.
This example shows that, in classical mechanics, the variables are in excess and that the discrete motion equation is also a conservation of mechanical energy.

Thus the continuous and discrete formulations give equivalent results with equations which are radically different. So what explains this concordance of the results? The classical Euler equation alone cannot solve this problem, it must necessarily be accompanied by the law of conservation of mass and a state law linking pressure, density and temperature, here the ideal gas law. These physical quantities are not all independent because the celerity is written here $ c_l = \sqrt{\gamma \: r \: T}$. The hyperbolic nature of the system of equations obtained is due to the temporal coupling between the Euler equation and the conservation of mass which leads to the appearance of the expected shocks on each of the variables.

Although various other numerical methods are more appropriate for capturing the progression of the various shocks in this flow, the discrete system (\ref{euler}) is capable of reproducing the theoretical solution of the problem to a relatively good degree of accuracy.

\subsection {Stokes paradox}

The explanations of certain artefacts of the equations of physics are manifold and do not always rest on objective reasons. The case of Stokes paradox is attributed to the dimensionality of space, the absence of inertial effects, turbulence, etc.
In two dimensions of space, the flow at constant velocity $V_0$ at infinity around a profile generates an artefact called Stokes paradox. For a circular section profile of radius R the solution of Stokes' equation can be found in polar coordinates $(r, \theta)$ using the incompressibility constraint $\nabla \cdot \bm v = 0$ and applying the curl operator. The boundary conditions correspond to the no-slip on the cylinder and a uniform velocity $V_0 = 1$ to infinity.
The solution which no longer depends on the viscosity $ \mu $ can be sought by setting $ \bm v = \nabla \times \mathbf \Psi $ where $\mathbf \Psi$ is the vector potential of the velocity which is reduced to the only component orthogonal to the plane $(r, \theta)$. Stokes' equation becomes $ \nabla^4 \psi = 0$ where $\psi (r, \theta)$ is the stream-function whose solution is written:
\begin{eqnarray}
\displaystyle{ \psi(r, \theta) = A \: \left( 2 \: r \: ln r - r + \frac{1}{r} \right) \: \sin \theta}
\label{fcour}
\end{eqnarray}

The resolution of the Stokes equation can be found in many textbooks of fluid mechanics \cite {Hap63}, \cite {Bat67}. This solution is only defined to a constant $A$ which must be zero if we apply the condition to infinity; this is what constitutes the Stokes paradox, there is no solution of this two-dimensional equation for a uniform flow at infinity. The long-distance solution must then be replaced by the Oseen solution. To suppress the paradox an asymptotic development method \cite {Van64} with a small parameter, the Reynolds number, can also be used.
The classic argument generally put forward to explain this phenomenon is that, at a great distance from the circle, the inertial effects cannot be neglected.

In discrete mechanics the viscous dissipation makes it possible to advance an explanation; this can be represented by the parameter $ - \sigma \: \bm v $ within the equation of motion:
\begin{eqnarray}
\displaystyle{  \nabla^4  \psi - \sigma \: \nabla^2  \psi = 0}
\label{stocal}
\end{eqnarray}

The solution on the stream-function (vector potential of the velocity) is then sought in the form $ \psi (r, \theta) = f (r) \: g (\theta) $; the vector potential of the acceleration $ \bm \psi = \nu \: \nabla \times \bm v $ is then calculated; they write:
\begin{eqnarray}
\left\{
\begin{array}{llllll}
\displaystyle{ \psi =  \left( r - \frac{a}{r} + b \: K_1( \sigma \: r) \right) \sin \theta  } \\ \\
\displaystyle{ \bm \psi = - 2 \: \nu \: \left(  1 - \frac{1}{r} + \frac{1}{r^2} \right) \: \frac{ K_1(\sigma r) }{  K_0(\sigma) } \: \sin \theta \: \mathbf e_z }
\end{array}
\right.
\label{mecasol}
\end{eqnarray}
with $ a = (\sigma \: K_0 (\sigma) + 2 \: K_1 (\sigma)) / (\sigma \: K_0 (\sigma)) $ and $ b = 2 / (\sigma \: K_0 ( \sigma)) $; $ K_0 $ and $ K_1 $ are the modified Bessel functions of order zero and one.

The Stokes paradox is thus removed, the velocity is zero on the circle and equal to $ V_0 = 1 $ at infinity. Whatever the value of $\sigma \ne 0$ the solution (\ref{mecasol}) allows to find a physical behavior.
In the discrete motion equation, viscous dissipation can be introduced (i) from the damping factor $ \alpha_t $ of the transverse waves (ii) by a negative acceleration $- \sigma \: \bm v$. In both cases the energy available in the flow is transformed into heat.

The Stokes equation or more generally the Navier-Stokes equation does not contain any viscous dissipation term; it can be calculated {\it a posteriori} in the form of a dissipation function $ \phi_d = \lambda (\nabla \cdot \bm v)^2 + \mu \: (\nabla \bm v + \nabla^t \bm v): \nabla \bm v$ within the energy equation; the use of this conservation law followed by the state law can lead to a loss of coherence on mechanical energy.

\subsection {Differences with the Navier-Lam{\'e} equation}

The computation of the constraints and displacements can be carried out by looking for the field of displacements $ \bm u $ or that of the constraints but this last method requires the satisfaction of compatibility conditions {\it i.e.} to ensure that the state of stress leads to a state of deformation compatible with the field of displacement by the intermediary of the constitutive law.

The most classic form of the Navier-Lam{\'e} equation, obtained by developing the divergence of the stress tensor $ \bm \sigma $, is written:
\begin{eqnarray}
\displaystyle{ \rho \: \frac{\partial^2 \bm u}{\partial t^2} = \left( \lambda + 2 \: \mu \right) \: \nabla \left( \nabla \cdot \bm u \right) - \mu \: \nabla \times \left( \nabla \times \bm u \right) + \mathbf f_s   }
\label{navierlame}
\end{eqnarray}
where $ \bm u $ is the displacement field, $ \lambda $ the compression modulus or first coefficient of Lam{\'e} and $\mu$ is the shear modulus.

This form is close to the equation (\ref {discrete}) but it differs from it on several points:
\begin {itemize} [label = \textcolor {blue} {\textbullet}]
\vspace {1.mm}
\item the coefficients $ \lambda $ and $ \mu $ are outside the operators which inhibits any possibility of showing the two components of a Helmholtz-Hodge decomposition;
\item $ M = \left (\lambda + 2 \: \mu \right)$ is only one and the same coefficient, the P-wave-modulus;
\item the two curls of the shear term are not of the same nature; the inner curl is applied to a polar vector and the outer curl is applied to a pseudo-vector. In discrete mechanics the primal and dual curls are expressed more clearly on the geometric topologies;
\item the equation (\ref{navierlame}) does not have terms of inertia, the term of the second order in time is not enough to describe the fast evolutions of the medium, for the vibrations at high frequencies for example;
\item the energies of compression and shearing do not appear explicitly within the equation what translates the intantaneous nature of the physical model, the incrementation of a state of mechanical equilibrium to another must be carried out of one another way.
\end {itemize}

Besides, the choice of the variable, the displacement $\bm u$, is not debatable but that of the velocity $\bm v$ such that $\bm u = \bm u^o + dt \: \bm v$ is just as appropriate. The reference state $\bm u^ o$ commonly used in solid mechanics becomes the current state of a temporal process where the equilibrium state at time $t$ is calculated from the knowledge of the one at the moment $t^o$. Like the discrete equation of motion, the Navier-Lam{\'e} equation is not associated with a law of conservation of mass, from this point of view it is autonomous.

\subsection{A formulation ready to use}

The equation of motion, Navier-Lam{\'e} for solids or Navier-Stokes for fluids, is local, expressed at a point in the inertial reference frame of the figure (\ref{primdual}a); Their resolution must be carried out starting from spatial and temporal discretizations. Many methods (Finite Differences, Finite Elements, Spectral methods, ...) and many schemes are developed to transform the continuous equation into a system of algebraic equations which leads to the resolution of a linear system. The location of the unknowns is also the subject of variants where all the unknowns are placed at a point (collocated grids) or broken down on the stencil (staggered grids) \cite{Har65}.

Discrete mechanics does not require this discretization step, the operators are immediately transposable on geometric polygonal or polyhedral topologies. The unknowns are the components of the velocity $\bm v$ on each segment $\Gamma$. The proposed concept is both a physical modeling of phenomena and a discretization of space.

The operators of the vector equation of discrete mechanics already have a geometric meaning in a three-dimensional space. These discrete operators can be defined simply from the basic geometric topology presented in figure (\ref{primdual}b). First, the discrete gradient is calculated as a difference, for example the scalar potential gradient $ \phi $ will be written $\nabla \phi = (\phi_b - \phi_a) / d$. It can be seen from the outset that the gradient vector is not that of continuum mechanics and represents a scalar on edge $ \Gamma $ oriented in the direction $\mathbf t$. The primal curl of vector $ \bm v $ is calculated as the circulation over all the edges of the oriented primal surface with $\nabla \times \bm v$ and will be carried by the unit vector $\mathbf n$ \cite{Cal19a}. The divergence of a vector, for example, is expressed at a point from the flows of the different oriented segments that converge towards it. The fourth operator is the dual curl $\nabla \times \bm \psi$ where the components of $ \bm \psi $ are orthogonal to the primal surfaces.
It should be noted that the 2D / 3D distinction does not exist. Indeed, even for a planar primal topology, the vector $\bm \psi$ is carried by the unit vector $\mathbf n$ orthogonal to this surface \cite{Cal20a}.

The two operators, gradient and dual curl, are those that project the action of different effects on the $\Gamma$ segment. This oriented edge is also the one on which the conservation of the acceleration will be carried out and where the various vector quantities will be evaluated, in particular the components $\bm v$ of the velocity.

In the selected topological structure, some operators are exact in the sense that the numerical error committed to evaluate them in a discrete point of view is zero. This is the case of the gradient which is defined by a difference and of the primal curl which is calculated from the Stokes theorem as the circulation of the vector on the contour $\Gamma$. The two other operators, divergence and dual curl, induce numerical errors that depend on the quality of the mesh used and the way in which the dual space is built; when the dual surfaces of the mesh are planar and that the line joining two barycentres is orthogonal to $\mathbf t$, these two operators are also exact and make it possible to obtain an exact solution when this one corresponds to a polynomial of degree inferior or equal to two.

Whereas classical mechanics has been established mainly by considering the divergence theorem for the relation between a flux on a surface and a volume and then making the elementary control volume tend to zero to obtain a formulation at a vertex, the discrete mechanics derive the equation of motion from the fundamental theorem of analysis and its consequences, {\it i.e.} the Stokes theorem in particular.
 If we denote the discrete quantities by the index $h$, two important properties of continuum, $\nabla_h \times \nabla_h \phi = 0$ and $\nabla_h \cdot \nabla_h \times \bm \psi = 0$ are verified in discrete mechanics. It is easy to show that, whatever the polygonal or polyhedral topologies, (i) the discrete curl of a discrete gradient is zero, (ii) the discrete divergence of curl calculated on the dual volume is nil.
In addition it should be noted that the decomposition of the acceleration into a divergence-free component and another to curl-free is globally orthogonal \cite{Cal20a} but also locally.

\section{Conclusions}

The Navier-Stokes and Navier-Lam{\'e} equations are well representative of fluid motions and movements in solids. In certain specific cases they present proven insufficiencies or paradoxes which limit their field of application and especially their unification. Two centuries after the appearance of the first works on theoretical aspects, the continuum mechanics presents, at least in their form, two different versions of the same problem. A good part of the difficulties noted are due to the primary concept of continuous medium, of the allocation of all the quantities at a point. A basic example, the inertia terms of the mechanics of continuous media, cannot be transformed into a Helmholtz-Hodge decomposition.

`` Why should we abandon well-established approaches such as 'the notion of continuous medium' in favor of another one? ''. Certainly, the continuum mechanics has definite advantages over point derivation, integration, analysis, but also leads to artifacts. For example, the use of mass conservation closely associated with the Navier-Stokes equation compensates for the lack of knowledge of the first Lam{\'e} coefficient for fluids. The association of conservation laws, constitutive equations, of state laws most often allows the inadequacy of the equation to be masked, but this is to the detriment of its consistency. Like Fourier's law for heat transfer, Fick's law for mass transfer, Newton's linear law is not sufficient to describe phenomena on all time scales.

This article only devoted to mechanics should not forget that many applications require sophisticated couplings with other fields of physics, electromagnetism, nonlinear optics, heat and mass transfers, etc. It is therefore legitimate to look for new ways to improve models of interaction.

\vspace{3.mm}

\textcolor{blue}{{\bf Author contributions} }

Author: Physical modeling, Conceptualization, Methodology, Research code,  Validation, Writing- Original draft preparation, Reviewing and Editing.

The paper has been checked by a proofreader of English origin.

\vspace{3.mm}

\textcolor{blue}{{\bf Declaration of Competing Interest} }

There are no conflict of interest in this work.

\vspace{3.mm}


\bibliographystyle{spmpsci}      


\begin{thebibliography}{}
\expandafter\ifx\csname url\endcsname\relax
  \def\url#1{\texttt{#1}}\fi
\expandafter\ifx\csname urlprefix\endcsname\relax\def\urlprefix{URL }\fi
\expandafter\ifx\csname href\endcsname\relax
  \def\href#1#2{#2} \def\path#1{#1}\fi

\end{thebibliography}


\begin{thebibliography}{10}
\providecommand{\url}[1]{{#1}}
\providecommand{\urlprefix}{URL }
\expandafter\ifx\csname urlstyle\endcsname\relax
  \providecommand{\doi}[1]{DOI~\discretionary{}{}{}#1}\else
  \providecommand{\doi}{DOI~\discretionary{}{}{}\begingroup
  \urlstyle{rm}\Url}\fi

\bibitem{Bat67}
Batchelor, G.: An Introduction to Fluid Mechanics.
\newblock Cambridge Univ. Press, Cambridge (1967)

\bibitem{Bur15}
Buresti, G.: A note on {S}tokes's hypothesis.
\newblock Acta Mechanica \textbf{226}, 3555--3559 (2015).
\newblock \doi{10.1007/s00707-015-1380-9}

\bibitem{Cal19a}
Caltagirone, J.P.: Discrete Mechanics, concepts and applications.
\newblock ISTE, John Wiley \& Sons, London (2019).
\newblock \doi{10.1002/9781119482826}

\bibitem{Cal19c}
Caltagirone, J.P.: Physique discr{\`e}te et relativit{\'e}.
\newblock Annales de la Fondation Louis de Broglie \textbf{44}, 1--13 (2019)

\bibitem{Cal20b}
Caltagirone, J.P.: Non-{F}ourier heat transfer at small scales of time and
  space.
\newblock International Journal of Heat and Mass Transfer \textbf{160}, 120145
  (2020).
\newblock \doi{10.1016/j.ijheatmasstransfer.2020.120145}

\bibitem{Cal20c}
Caltagirone, J.P.: On {H}elmholtz-{H}odge decomposition of inertia on a
  discrete local frame of reference.
\newblock Phys. Fluids \textbf{32}, 083604 (2020).
\newblock \doi{10.1063/5.0015837}

\bibitem{Cal21b}
Caltagirone, J.P.: Application of discrete mechanics model to jump conditions
  in two-phase flows.
\newblock J. Comp. Physics \textbf{432}, 110151 (2021).
\newblock \doi{10.1016/j.jcp.2021.110151}

\bibitem{Cal21d}
Caltagirone, J.P.: On a reformulation of {N}avier-{S}tokes equations based on
  {H}elmholtz-{H}odge decomposition.
\newblock Phyics of Fluids \textbf{33}, 063605 (2021).
\newblock \doi{10.1063/5.0053412}

\bibitem{Cal20a}
Caltagirone, J.P., Vincent, S.: On primitive formulation in fluid mechanics and
  fluid-structure interaction with constant piecewise properties in
  velocity-potentials of acceleration.
\newblock Acta Mechanica \textbf{231}(6), 2155--2171 (2020).
\newblock \doi{10.1007/s00707-020-02630-w}

\bibitem{Des05}
Desbrun, M., Hirani, A., Leok, M., Marsden, J.: Discrete exterior calculus.
\newblock arXiv/math/0508341v2 pp. 1--53 (2005)

\bibitem{Gad95b}
Gad-El-Hak, M.: Stokes hypothesis for a newtonian, isotropic fluid.
\newblock J. of Fluids Engineering \textbf{117}(1), 3--5 (1995).
\newblock \doi{10.1115/1.2816816}

\bibitem{Gue06}
Guermond, J., Minev, P., Shen, J.: An overview of projection methods for
  incompressible flows.
\newblock Comput. Methods Appl. Mech. Engrg. \textbf{195}, 6011--6045 (2006).
\newblock \doi{10.1016/j.cma.2005.10.010}

\bibitem{Ham08}
Hamman, C., Klewick, J., Kirby, R.: On the {L}amb vector divergence in
  {N}avier-{S}tokes flows.
\newblock J. Fluid Mech. \textbf{610}, 261--284 (2008).
\newblock \doi{10.1017/S0022112008002760}

\bibitem{Hap63}
Happel, J., H.~Brenner, H.: Low Reynolds Number Hydrodynamics.
\newblock Kluwer Academic Publishers, Boston (1963)

\bibitem{Har65}
Harlow, F., Welch, J.: Numerical calculation of time-dependent viscous
  incompressible flow of fluid with a free surface.
\newblock Physics of Fluids \textbf{8}, 2182--2189 (1965).
\newblock \doi{10.1063/1.1761178}

\bibitem{Hol11}
Holmes, M., Parker, N., Povey, M.: Temperature dependence of bulk viscosity in
  water using acoustic spectroscopy.
\newblock J. Phys. Conf. series \textbf{269} (2011).
\newblock \doi{10.1088/1742-6596/269/1/012011}

\bibitem{Noe11}
Kosmann-Schwarzbach, Y.: Noether {T}heorems. Invariance and {C}onservations
  {L}aws in the {T}wentieth {C}entury.
\newblock Springer-Verlag, New York (2011).
\newblock \doi{10.1007/978-0-387-87868-3}

\bibitem{Lan59}
Landau, L., Lifchitz, E.: Fluid Mechanics.
\newblock Pergamon Press, London (1959)

\bibitem{Lan71b}
Landau, L., Lifchitz, E.: The Classical Theory of Fields, Third Revised English
  Edition.
\newblock Pergamon Press Ltd, Oxford (1971)

\bibitem{Raj13}
Rajagopal, K.: A new development and interpretation of the {N}avier-{S}tokes
  fluid which reveals why the "{S}tokes assumption" is inapt.
\newblock International Journalof Non-Linear Mechanics \textbf{50}, 141--151
  (2013).
\newblock \doi{10.1016/j.ijnonlinmec.2012.10.007}

\bibitem{Sha96}
Shaskov, M.: Conservative Finite-Difference Methods on General Grids.
\newblock Boca Raton: CRC Press (1996).
\newblock \doi{10.1201/9781315140209}

\bibitem{Sod78}
Sod, G.: A survey of several finite difference methods for systems of nonlinear
  hyperbolic conservation laws.
\newblock J. Comput. Phys. \textbf{27}, 1--31 (1978).
\newblock \doi{10.1016/0021-9991(78)90023-2}

\bibitem{Tru54}
Truesdell, C., Rosenhead, L.: The present status of the controversy regarding
  the bulk viscosity of fluids.
\newblock Proc. R. Soc. Lond. A \textbf{226} (1954).
\newblock \doi{10.1098/rspa.1954.0237}

\bibitem{Van64}
Van~Dyke, M.: Perturbation Methods in Fluids Mechanics.
\newblock Academic Press, California University (1964)

\bibitem{Yua13}
Yuan, G., Jiulin, S., Kaixing, Z., Xindao, H.: Determination of bulk viscosity
  of liquid water via pulse duration measurements in stimulated brillouin
  scattering.
\newblock Chinese Optics Letters \textbf{11} (2013).
\newblock \doi{10.3788/COL201311.112902}

\end{thebibliography}

\end{document}